# Wavelet Transform-Based Analysis of QRS complex in ECG Signals

Swapnil Barmase, Saurav Das, Sabyasachi Mukhopadhyay

Indian Institute of Science Education and Research, Kolkata, Mohanpur campus, Nadia 741252

*Abstract:* In the present paper we have reported a wavelet based time-frequency multiresolution analysis of an ECG signal. The ECG (electrocardiogram), which records hearts electrical activity, is able to provide with useful information about the type of Cardiac disorders suffered by the patient depending upon the deviations from normal ECG signal pattern. We have plotted the coefficients of continuous wavelet transform using Morlet wavelet. We used different ECG signal available at MIT-BIH database and performed a comparative study. We demonstrated that the coefficient at a particular scale represents the presence of QRS signal very efficiently irrespective of the type or intensity of noise, presence of unusually high amplitude of peaks other than QRS peaks and Base line drift errors. We believe that the current studies can enlighten the path towards development of very lucid and time efficient algorithms for identifying and representing the QRS complexes that can be done with normal computers and processors.

*Keywords*- ECG signal, Continuous Wavelet Transform, Morlet Wavelet, Scalogram, QRS Detector.

1. Introduction

The ECG (electrocardiogram), which is nothing but recording hearts electrical activity, provides very vital information about the wide range of Cardiac disorders depending upon the deviations from normal ECG signal pattern. Biometric signal analysis has attained increased attention of researchers because of the direct implications corresponding to patients conditions by detecting abrupt changes in patterns of signal. In general, the frequency range of an ECG signal varies between 0.05–100 Hz with the dynamic range between 1–10 mV. The ECG signal is characterized by five peaks and valleys labeled by the letters P, Q, R, S, T. In some cases we also use another peak called U.The QRS complex is the most prominent waveform
within the electrocardiographic (ECG) signal, with normal duration from 0.06 s to 0.1 s.[1] The performance of ECG analyzing system depends mainly on the accurate and reliable detection of the QRS complex, as well as T- and P waves.However since the QRS complexes have a time-varying morphology, they are not always the strongest signal component in an ECG signal. In addition there are many sources of noise in a clinical environment, for example, power line interference, muscle contraction noise, poor electrode contact, patient movement, and baseline wandering due to respiration that can degrade the ECG signal.

The algorithms for QRS detectors can be classified into three categories: 1) nonsyntactic, which are time consuming owing to the need forgrammar inference for each class of patterns[2]. 2) syntactic,[3,4,5] which are used widely and 3) hybrid.Previously applied algorithms commonly use nonlinear filtering to detect QRS complexes using thresholding,[4] artificial intelligence using hidden Markov models,[6] and timerecursive prediction techniques[7] A General algorithm is passing the signal passed through a nonlinear transformer like derivative and square, etc., toenhance the QRS complexes after filtering the ECG signal using a bandpass filter to suppress the *P* and *T* waves and noise and finally determining the presence of QRS complexes using decision. The main drawbacks of these techniques is that frequency variation in QRS complexes adversely affects their performance. The frequency band of QRS complexes generally overlaps the frequency band of noise, resulting in both false positives and false negatives. Methods using artificial intelligence are time consuming due to the use of grammar and inference rules as mentioned earlier.[2]The hidden Markov model approach too requires considerable time even with the use of efficient algorithms. Wavelet analysis is a very promising mathematical tool 'a mathematical microscope' that gives good estimation of time and frequency localization. Wavelet analysis has become a renowned tool for characterizing ECG signal and some very efficient algorithms has been reported using wavelet transform as QRS detectors.[8,9,10] Tai et. Al. has reported that their algorithm has the detection rate of QRS complexes above 99.8% for the MIT/BIH database, using quadratic spline wavelet, however the algorithm along with being complex cannot be run in normal computers.
In this paper we have reported methodologies that are very simple in order to develop algorithms to detect the QRS complex using continuous wavelet transform. We used the Morlet wave and plotted coefficients to detect the QRS peaks with almost full efficiency with respect to the signal that we processed having various types and intensities of noise, unusually high amplitude of peaks other than QRS and base line drifts.

## 2. Theory
### 2.1 Continuous Wavelet transform:

The wavelet transform of a continuous time signal, x(t), is defined as: $T(a,b) = \frac{1}{\sqrt{a}} \int_{-\infty}^{\infty} x(t) \psi^*(\frac{t-b}{a}) dt$, where $\psi^*(t)$ is the complex conjugate of the wavelet function of $\psi(t)$, $a$ is the dilation parameter of the wavelet and $b$ is the location parameter of the wavelet. In order to be classified as a wavelet, the function must satisfy certain mathematical criteria. These are:

A wavelet must have finite energy: $E = \int_{-\infty}^{\infty} |\psi(t)|^2 dt < \infty$.

If $\psi(f)$ is the fourier transform of $\psi(t)$,

$$\psi(\omega) = \int_{-\infty}^{\infty} \psi(t) e^{-j\omega t} dt$$

Then the following condition must hold: $C_g = \int_0^{\infty} \frac{|d\omega|^2}{\omega} d\omega < \infty$ ……………………………………….(1)

This implies that the wavelet has no zero frequency component, i.e, $\psi(0) = 0$, or to put it another way, it must have a zero mean. Equation (1) is known as the 'Admissibility Condition' and $C_g$ is called 'Admissible Constant'.

The value of $C_g$ depends on the chosen wavelet.

For complex (or analytic) wavelets, the Fourier transform must both be real and vanish for negative frequencies.

The contribution to the signal energy at the specific $a$ scale and $b$ location is given by the two-dimensional wavelet energy density function known as the 'Scalogram':

$E(a,b) = |T(a,b)|^2$.

The total energy in the signal may be found from its wavelet transform as follows:

$$E = \frac{1}{C_g} \int_{-\infty}^{\infty} \int_0^{\infty} \frac{1}{a^2} |T(a,b)|^2 da db$$

In practice a fine discretisation of the continuous wavelet transform is computed where usually the $b$ location is discretised at the sampling interval and the $a$ scale is discretised logarithmically. The $a$ scale discretisation is often taken as integer powers of 2, however, we use a finer resolution in our method where the $a$ scale scale discretisation is in fractional powers of two. This discretisation of the continuous wavelet transform (CWT) is made distinct from the discrete wavelet transform (DWT) in the literature. In its basic form, the DWT employs a dyadic grid (integer power of two scaling in $a$ and $b$) and orthonormal wavelet basis functions and exhibits zero redundancy. Our method, i.e. using a high resolution in wavelet space as described above, allows individual maxima to be followed accurately across scales, something that is often very difficult with discrete orthogonal or dyadic stationary wavelet transforms incorporating integer power of two scale discretisation.

## 3. Results and Discussions

The data has been taken from MIT-BIH arrhythmia database. We analyzed different signal of length 10 seconds for our algorithm and analysis have some different types of deviations from normal specifically. Namely, Record 105, which is more noisy than the others; Record 108 has unusually high and sharp P waves; Record 203 has a great number of QRS complexes with multiform ventricular arrhythmia; and Record 222 has some non-QRS waves with highly unusual morphologies and 109 having an base line drift in the signal which is one of the major problem causing failure of threshold type algorithms. As shown in figures 1 to 10 we have analyzed different signal with different types of noises, errors and fluctuations and we see that in the coefficient plot we get a band of high energy in the scale range of 220 to 250 corresponding to exact number of the QRS peaks available in the signal. Very efficient and lucid algorithms can be developed to read this plot at a particular scale. We have done a localized analysis of a signal for a particular duration and one can count manually to check the results. For a very long duration signal, a variable scale can be defined to make the perfect count of QRS signal.

Another inference that can be drawn from the Coefficients plot is about the exact positions where the energy scale representation is zoomed by the continuous wavelet transform.

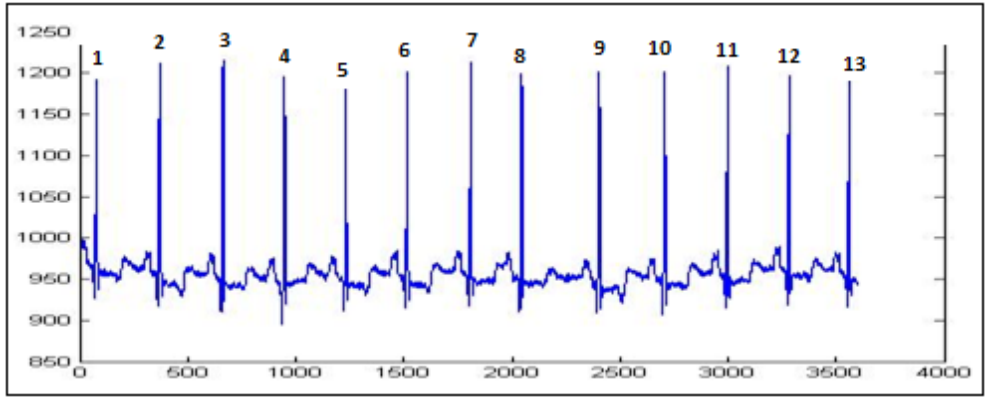

**Figure 1: Analyzed signal of record 100 from MIT-BIH database**

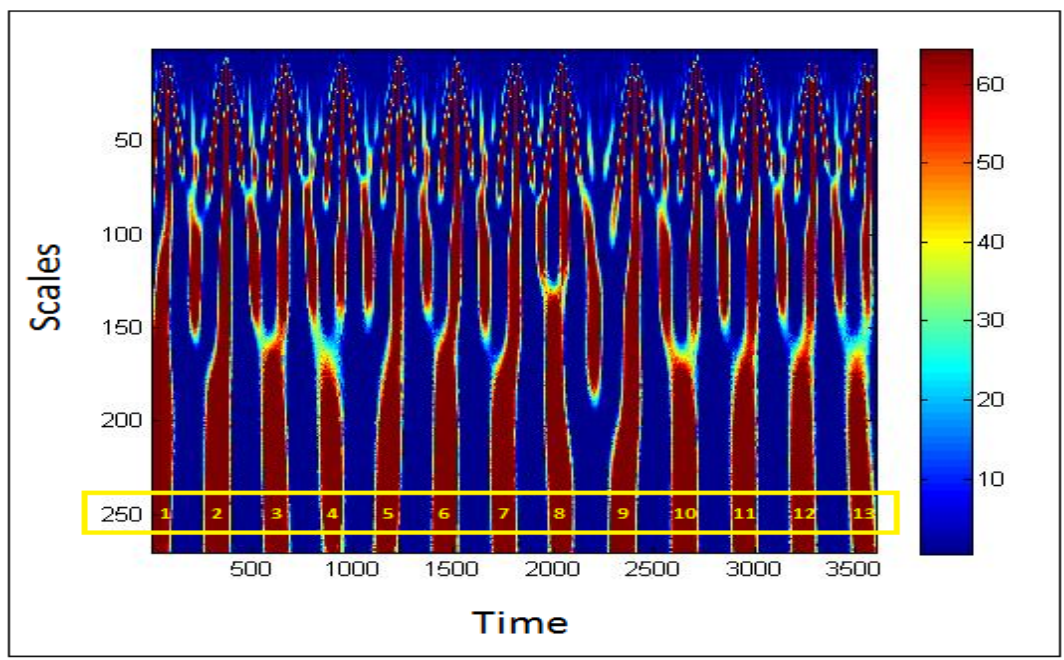

**Figure 2: Coefficient plot after applying Morlet wavelet**

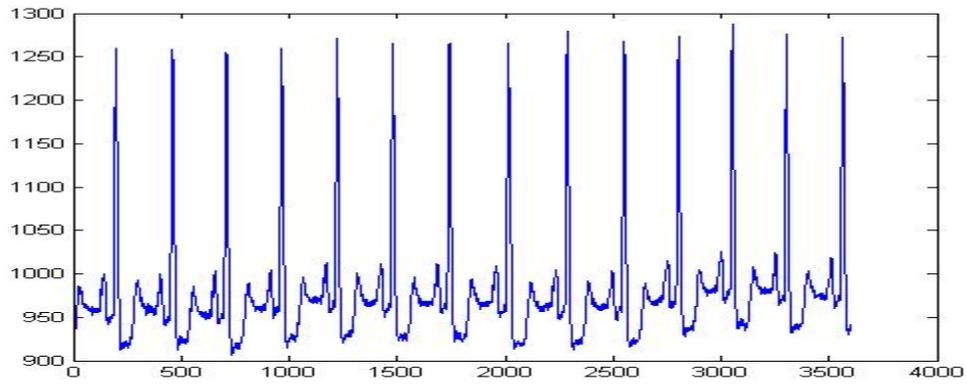

Figure 3: Analyzed signal, record 105 having unusually high noise

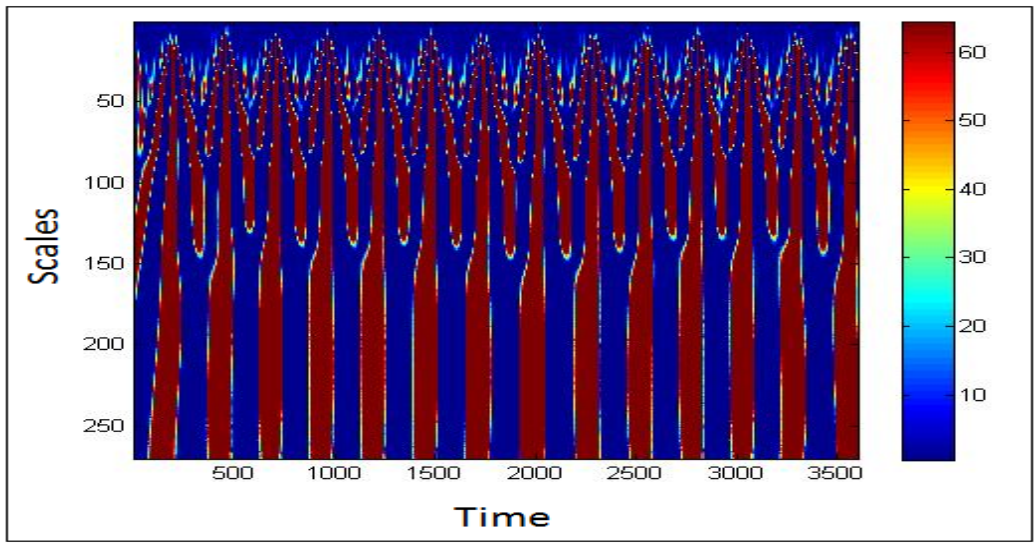

Figure 4: Coefficients plot of MIT-BIH record 105 ECG signal

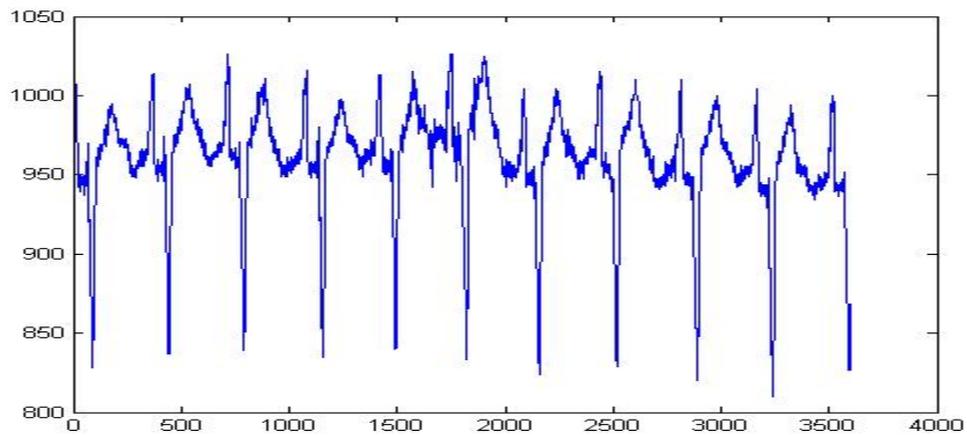

Figure 5: Analyzed signal, record 108 having unusually high amplitude of P-waveforms

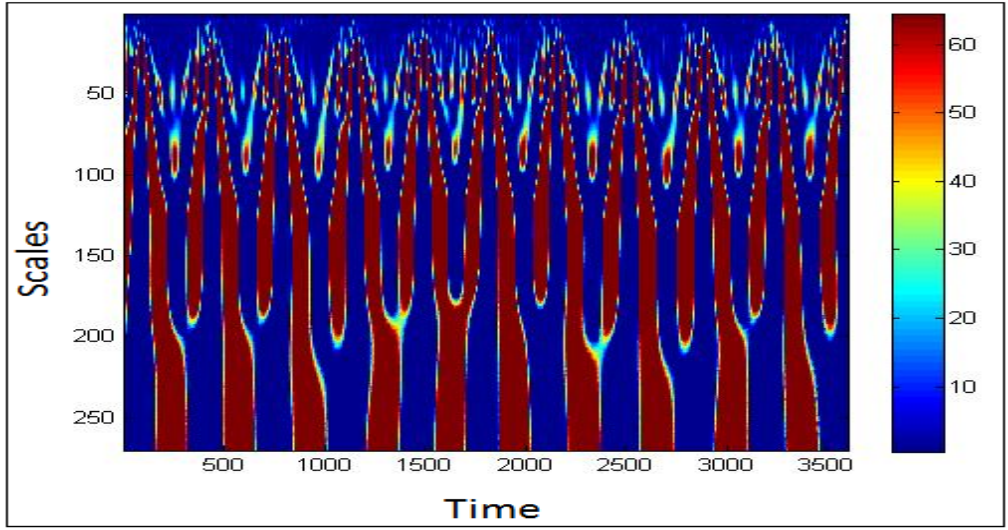

Figure 6: Coefficients plot of MIT-BIH record 108 ECG signal

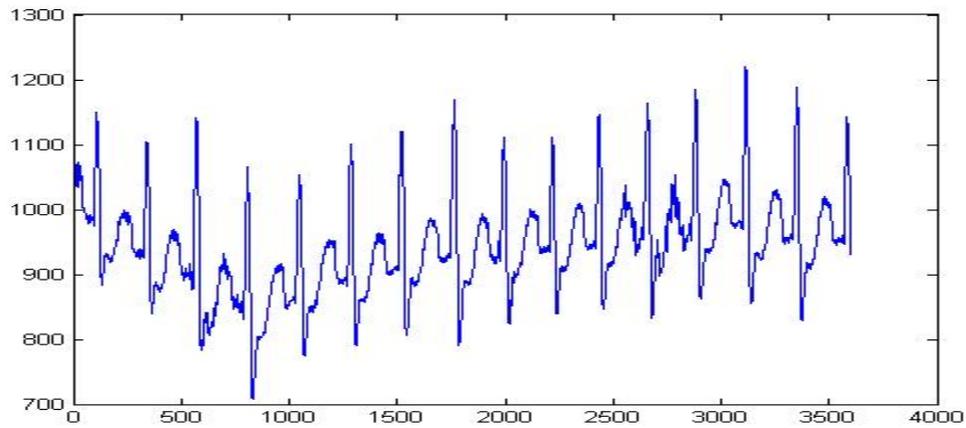

Figure 7: Analyzed signal, record 109 having a shifting baseline, an error called as base line drift

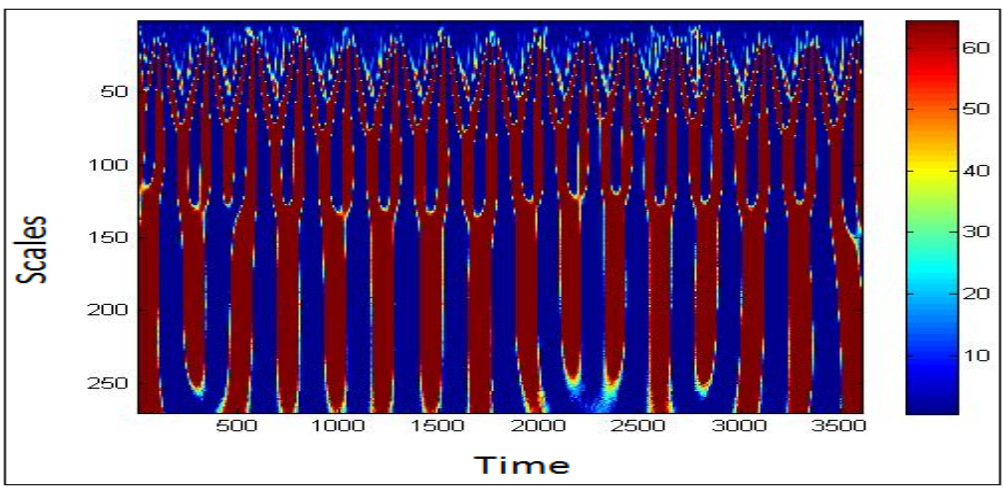

Figure 8: Coefficients plot of MIT-BIH record 109 ECG signal

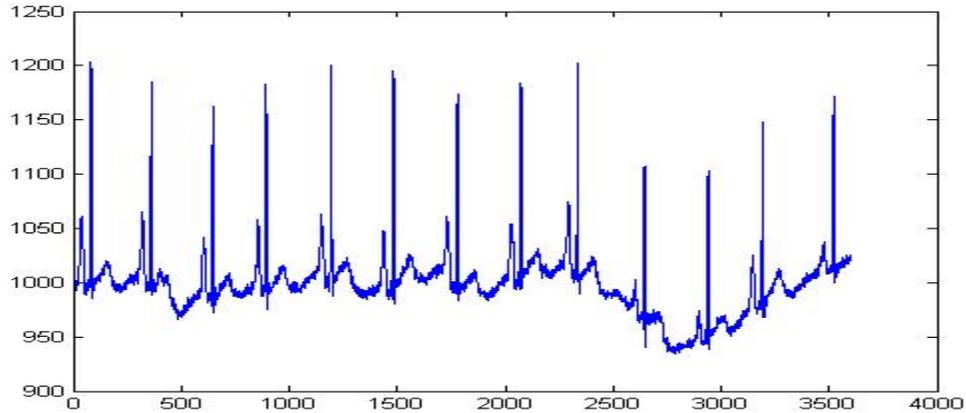

Figure 9: Analyzed Signal, MIT-BIH record 222 has some non-QRS waves with highly unusual morphologies

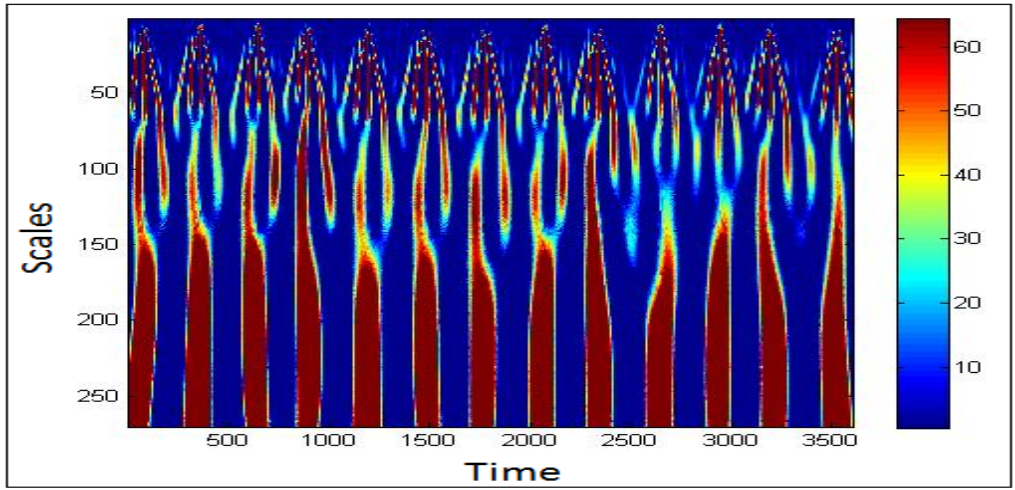

Figure 10: Coefficients plot of MIT-BIH record 222 ECG signal

4. **Conclusion:**

In the present paper we have reported a time-frequency multiresolution analysis of an ECG signal. We have plotted the coefficients of continuous wavelet transform using Morlet wavelet. We used different ECG signal available at MIT-BIH database and performed a comparative study. We demonstrated that the coefficient at a particular scale represents the presence of QRS signal very efficiently irrespective of the type or intensity of noise, presence of unusually high amplitude of peaks other than QRS peaks and Base line drift errors. We believe that the current studies can enlighten the path towards development of very lucid and time efficient algorithms for identifying and representing the QRS complexes that can be done with normal computers and processors. In addendum we suggest that with few modifications of the current work can reveal the features and characteristics of other ECG waveform viz. P and T waveform which can also provide with some important information about physiological conditions of patient suffering from heart disease.